\documentclass[twocolumn]{aastex62}

\usepackage{amsmath}

\newcommand\orion{\texttt{ORION2}}
\newcommand\pluto{\texttt{PLUTO}}
\newcommand\chianti{\texttt{CHIANTI}}
\def\isoRun{\texttt{IsoRun}}
\def\adiaRun{\texttt{AdiaRun}}

\def\hii{H~\textsc{ii}}
\def\msun{M_{\rm \odot}}
\def\lsun{L_{\rm \odot}}
\def\rsun{R_{\rm \odot}}
\def\teff{T_{\rm eff}}
\def\vinf{v_{\rm \infty}}

\usepackage[normalem]{ulem}

\received{\today}
\revised{xxx}
\accepted{xxx}
\submitjournal{ApJ}

\shorttitle{Wind Bubbles Around Stars}
\shortauthors{Rosen et al.}

\begin{document}

\title{Blowing Bubbles around Intermediate-Mass Stars: Feedback from Main-Sequence Winds is not Enough}

\correspondingauthor{Anna Rosen}
\email{anna.rosen@cfa.harvard.edu}

\author[0000-0003-4423-0660]{Anna L. Rosen}
\altaffiliation{NASA Einstein Fellow}
\altaffiliation{ITC Fellow}
\affiliation{Center for Astrophysics $|$ Harvard \& Smithsonian, 60 Garden St, Cambridge, MA 02138, USA}

\author[0000-0003-1252-9916]{Stella S. R. Offner}
\affiliation{Department of Astronomy, The University of Texas at Austin, 2500 Speedway, Austin, TX 78712, USA}

\author[0000-0002-6747-2745]{Michael M. Foley}
\affiliation{Center for Astrophysics $|$ Harvard \& Smithsonian, 60 Garden St, Cambridge, MA 02138, USA}

\author[0000-0002-1790-3148]{Laura A. Lopez}
\affiliation{Department of Astronomy, The Ohio State University, 140 W. 18th Ave., Columbus, OH 43210, USA}
\affiliation{Center for Cosmology and AstroParticle Physics, The Ohio State University, 191 W. Woodruff Ave., Columbus, OH 43210, USA}
\affiliation{Niels Bohr Institute, University of Copenhagen, Blegdamsvej 17, 2100 Copenhagen, Denmark}

\begin{abstract}
Numerous spherical ``shells" have been observed in young star-forming environments that host low- and intermediate-mass stars. These observations suggest that these shells may be produced by isotropic stellar wind feedback from young main-sequence stars. However, the driving mechanism for these shells remains uncertain because the momentum injected by winds is too low to explain their sizes and dynamics due to their low mass-loss rates. However, these studies neglect how the wind kinetic energy is transferred to the ISM and instead assume it is instantly lost via radiation, suggesting that these shells are momentum-driven. Intermediate-mass stars have fast ($v_w \gtrsim 1000$ km/s) stellar winds and therefore the energy injected by winds should produce energy-driven adiabatic wind bubbles that are larger than momentum-driven wind bubbles. Here, we explore if energy-driven wind feedback can produce the observed shells by performing a series of 3D magneto-hydrodynamic simulations of wind feedback from intermediate-mass and high-mass stars that are placed in a magnetized, turbulent molecular cloud. We find that, for the high-mass stars modeled, energy-driven wind feedback produces $\sim$pc scale wind bubbles in molecular clouds that agree with the observed shell sizes but winds from intermediate-mass stars can not produce similar shells because of their lower mass-loss rates and velocities. Therefore, such shells must be driven by other feedback processes inherent to low- and intermediate-mass star formation.

\end{abstract}


\section{Introduction} 
\label{sec:intro}

Stellar feedback -- the injection of energy and momentum by young stars -- heats up and shapes the interstellar medium (ISM), thereby making star formation inefficient in giant molecular clouds (GMCs) and star-forming galaxies \citep[see reviews by][]{Krumholz2019a, Rosen2020a}. Stars form from the gravitational collapse of dense molecular gas within GMCs and during their formation they inject energy and momentum into the ISM via radiation, magnetically launched collimated outflows, and isotropic stellar winds that are either driven by magnetic processes or by radiation pressure \citep[e.g., ][]{Matzner2000a, Cranmer2008a, Krumholz2009b, Offner2009a, Vink2001a, Rosen2014a, Rosen2020b, Gusjevnov2021a, Grudic2021a, Olivier2021a}. These feedback processes generate small-scale motions in the ISM, thereby driving turbulence in molecular clouds, which increases the lifetime of molecular clouds and possibly triggers subsequent star formation \citep{Arce2011a, Lili2015a, Offner2018a, Feddersen2018a, GallegosGarcia2020a, Xu2020a, Xu2020b, Menon2021a, Paron2021a}. 

Significant attention has been given to studying stellar feedback from high-mass stars ($\gtrsim 8 \, \msun$) since they dominate the energetics of the ISM in star-forming galaxies \citep{Lopez2011a, Lopez2014a, Rosen2016a, Grudic2019a, Rosen2020a, Barnes2020a, Olivier2021a}. Hundreds of mid-infrared and infrared bright bubbles with radii up to $\sim$10s pc have been identified in the Galactic plane with the Spitzer-GLIMPSE and WISE surveys \citep{Deharveng2010a, Anderson2014a}. Many of these bubbles surround \hii\ regions that are driven by photo-ionizing radiation and stellar wind feedback from young, massive stars, and the expanding swept-up dense shells may be potential sites of triggered star formation \citep[e.g.,][]{Brand2011a, Chauhan2011a, Zhou2020a}. In addition, numerous sub-pc to pc scale spherical shells, observed via CO emission, have been identified around intermediate-mass main sequence (MS) and pre-main sequence (PMS) stars (late-type B stars) in galactic star-forming regions, such as Perseus and Orion \citep{Arce2011a, Lili2015a, Feddersen2018a}. These studies concluded that the identified shells or ``bubbles" are likely produced by isotropic stellar feedback from young individual intermediate-mass stars or young multiple systems since they are roughly spherical. 
However, in many cases the stellar source creating the feature is not clear.  Taurus, for example, contains no B-type stars, so any shells there must be produced by lower mass stars \citep{Lili2015a,Xu2020a}. This presents a puzzle, since lower mass stars experience relatively low mass-loss rates as MS or PMS stars. Consequently,
the feedback processes driving such shells around young, intermediate-mass stars remain uncertain. 

By considering momentum-driven feedback from isotropic stellar winds (i.e., momentum-conserving), in which the wind kinetic energy is instantly lost via radiation (i.e., the ``slow-wind" model presented in \citet{Koo1992a}), these studies concluded that winds with mass-loss rates of $\dot{M}_{\rm w}\sim10^{-7}-10^{-6} M_{\rm \odot} \; \rm yr^{-1}$ and speeds of $v_{\rm w} = 200 \; \rm km/s$ can explain the structure and dynamics of the observed bubbles. Following these results, \citet{Offner2015a} modeled the impact of stellar wind feedback in a turbulent molecular cloud with magnetohydrodynamic (MHD) simulations using an isothermal equation of state, which assumes the kinetic energy from stellar winds is efficiently radiated away and is therefore not transferred to the ISM, with similar mass-loss rates and wind velocities predicted by \cite{Arce2011a}. Their results agreed with the wind bubble radii and dynamics observed in Perseus. 

While these inferred wind mass-loss rates are able to produce the observed bubbles that surround intermediate mass stars in star-forming environments, they are orders of magnitude larger than the expected $\dot{M}_{\rm w}\sim10^{-11}-10^{-9} M_{\rm \odot} \; \rm yr^{-1}$ values inferred from theoretical and observational studies of radiatively-driven winds from intermediate-mass stars \citep[e.g.,][]{Leitherer1992a, Vink2001a, Krticka2014a}. The high mass-loss rates derived from the shell kinematics imply that the driving sources are massive O-type stars, which are not present.
Additionally, radiatively-driven winds from intermediate-mass and massive stars have velocities of $v_{\rm w} \sim 1000$ km~s$^{-1}$ since they must leave the stellar surface at or above the escape speed of the star \citep{Leitherer1992a, Vink2001a}. These fast stellar winds will collide with the surrounding ISM and thermalize producing hot, shocked stellar wind material \citep{Castor1975a, Weaver1977a}. If the wind kinetic energy is fully thermalized, such that
\begin{equation}
\frac{\dot{E}_{\rm w} t_w}{V} = \frac{3}{2}n_{\rm X} k_{\rm B} T_{\rm X}
\end{equation}
\noindent
where $\dot{E}_{\rm w} = (1/2) \dot{M}_{\rm w} v_{\rm w}^2$ is the rate of kinetic energy injection by winds, $n_{\rm X} =  \dot{M}_{\rm w} t/(\mu m_{\rm p} V)$ is the density of the shock-heated wind material with mean molecular weight $\mu=0.6$ for a fully ionized plasma of solar composition that occupies volume $V$ and is injected for time $t_{\rm w}$, then the resulting thermalized wind material temperature is
\begin{equation}
T_{\rm X}  \approx 2.42 \times 10^7 \left(\frac{v_{\rm w}}{1000 \; \rm km/s}\right)^2 \; \rm K .
\end{equation}
The resulting hot gas will cool via adiabatic expansion rather than significant radiative losses since cooling is inefficient at these high temperatures resulting in energy-driven (i.e., energy-conserving) stellar wind feedback \citep{Castor1975a, Weaver1977a, Koo1992a, Rogers2013a, Rosen2014a}. The resulting expansion will sweep up a dense spherical shell of entrained molecular material, producing energy-driven adiabatic wind bubbles with larger radii than those expected from momentum-driven feedback alone (i.e., radiative wind bubbles).

Whether stellar wind feedback is energy-driven or momentum-driven remains highly debated. Observations of the hot gas generated by stellar wind feedback in star clusters, which host numerous massive stars, have shown that much of the wind kinetic energy injected by massive stars is missing \citep{Harper-Clark2009a, Lopez2011a, Lopez2014a, Rosen2014a}. Recent work by \citet{Lancaster2021a, Lancaster2021b} find that the missing energy can be accounted for by turbulent mixing and efficient cooling at the interface between the hot bubble driven by stellar wind feedback and the surrounding turbulent and cold ISM. In this scenario, the majority of the kinetic energy from stellar winds is lost via radiative cooling because the hot shock-heated gas and cold interstellar gas mix to form $\sim10^4$ K gas that cools efficiently. These studies find that the dynamics of gas that surrounds star clusters is influenced by momentum-driven stellar wind feedback, because the wind kinetic energy is radiated away. However, these studies do not take into account magnetic fields nor do they focus on wind feedback from individual stars, which we explore here.

In this paper, we investigate how momentum- and energy-driven wind feedback from radiatively-driven stellar winds can produce bubbles around intermediate- and high-mass stars by performing 3D MHD numerical simulations of radiatively-driven stellar wind feedback for young early-type (high-mass; $M_{\rm \star} \gtrsim 8 \msun$) and late-type (intermediate-mass; $M_{\rm \star} \gtrsim 3 \msun$) B stars in a turbulent and magnetized molecular cloud. This paper is organized as follows: we provide our theoretical motivation as to why winds from intermediate- and high-mass stars drive energy-conserving adiabatic wind bubbles in Section~\ref{sec:theory}; we describe our numerical methodology and simulation design in Section~\ref{sec:methods}; we present and discuss our results in Sections~\ref{sec:results} and \ref{sec:discussion}, respectively; finally, we conclude in Section~\ref{sec:conclusions}.

\section{Theoretical Motivation}
\label{sec:theory}
\subsection{Momentum- versus Energy-driven Wind Bubbles}
\label{sec:bubble}
\citet{Koo1992a} describe the structure and dynamics of  momentum-driven (i.e., fully radiative conditions in which the shocked wind material loses energy via efficient radiative losses) and energy-driven (i.e., non-radiative) wind bubbles produced by a star. They find that there exists a critical wind velocity, $v_{\rm cr}$, that differentiates between these two types of wind bubbles. If the wind velocity is below $v_{\rm cr}$ then the wind bubble is fully radiative and  therefore momentum-conserving (i.e., their ``slow winds" case) and for wind velocities much higher than this critical velocity (i.e., their ``fast winds" case) the cooling time is much longer than the age of the bubble and therefore the wind bubble does not experience significant radiative losses and is therefore adiabatic. In this ``fast winds" scenario the bubble shell dynamics are energy-driven. We summarize their derivation for $v_{\rm cr}$ here to motivate that main sequence winds from intermediate- and high-mass stars belong to the ``fast winds" scenario and therefore drive adiabatic energy-conserving wind bubbles.

Consider a star with mass-loss rate $\dot{M}_{\rm w}$ and constant wind velocity $v_{\rm w}$ that is highly supersonic. Initially, the mass of the wind injected is much larger than the mass of the swept-up ambient medium and so the wind material expands freely. This is referred to as the ``free expansion phase." At the outermost part of the wind, there are two shocks. The first is the ``ambient shock," which propagates into the ambient medium and the second is a reverse shock (i.e., the wind shock), which propagates back into the wind. The shocked ambient medium and the shocked wind material are separated by a  contact discontinuity. As the reverse shock rapidly reaches the center, the bubble interior is heated to very high temperatures. Initially, the wind shock is radiative since the wind density is sufficiently high so that cooling is efficient, but as the hot wind bubble adiabatically expands the pre-shock density decreases and its velocity increases so that the wind shock eventually becomes non-radiative. If $v_{\rm w}$ is sufficiently high (i.e., the ``fast winds" scenario) this transition occurs early while the wind is freely expanding such that the wind kinetic energy is roughly conserved and thermalizes the interior gas to very high temperatures, which does not rapidly cool via radiation. 

During this phase, the wind material expands almost freely up to radius $R_{\rm f}$ until the wind density is comparable to the ambient medium density, $\rho_0$:

\begin{equation}
    \rho_{w}(R_{\rm f}) = \frac{\dot{M}_{\rm w}}{4 \pi R^2_{\rm f} v_{\rm w}} \equiv \rho_0.
\end{equation}
\noindent
The corresponding timescale for the wind bubble radius to reach $R_{\rm f}$ is $t_{\rm f} = R_{\rm f}/v_{\rm w}$:
\begin{equation}
\label{eqn:tf}
\begin{split}
t_{\rm f} &\equiv \left(\frac{\dot{M}_{\rm w}}{4 \pi \rho_0 v_{\rm w}^3}\right)^{1/2} \\
	&= 0.046 \left(\frac{\dot{M}_{\rm w}}{10^{-9} \; \rm{\msun \; yr^{-1}}}\right)^{1/2} \left(\frac{n_0}{10^3 \; \rm cm^{-3}}\right)^{-1/2} \\
	& \hspace{1.25cm} \left(\frac{v_{\rm w}}{10^3 \; \rm{km \; s^{-1}}}\right)^{-3/2} \; \rm yr,
\end{split}
\end{equation}
\noindent
where we have taken $n_0 = \rho_0/\mu_{\rm H}$ where $\mu_{\rm H} = 2.34 \times 10^{-24} \; \rm g$ is the mean molecular weight for Hydrogen nuclei for a gas of cosmic abundances. \citet{Koo1992a} define the transition from the ``slow winds" case to the ``fast winds" case when $t_{\rm f}(v_{\rm cr}) = t_{\rm cool}(v_{\rm cr})$, where $t_{\rm cool}$ is the timescale at which the shock heated material efficiently cools via radiation. If $v_{\rm w} \gg v_{\rm cr}$ then the hot gas within the bubble does not experience significant radiative losses and therefore will evolve adiabatically.

\citet{Koo1992a} approximate the cooling time-scale for a parcel of gas with pre-shock density, $\rho_{\rm ps}$, engulfed by a shock moving at velocity $v_{\rm s}$ as $t_{\rm cool} = C_{1} v_{\rm s}^3/\rho_{\rm ps}$ where $C_{1}=6.0 \times 10^{-35} \rm \; g \; cm^{-6} \; s^4$ for a gas in ionization equilibrium assuming the cooling function, $\Lambda(T)$, is proportional to $T^{1/2}$. When the location of the wind shock is equal to $R_{\rm f}$ the shock velocity is roughly the wind velocity and the pre-shock density is equal to the ambient density. Therefore, the cooling time scale can be approximated as 

\begin{equation}
\label{eqn:tcool}
\begin{split}
t_{\rm cool} &\approx \frac{C_{1} v_{\rm w}^3}{\rho_0} \\
	&= 812 \left(\frac{n_0}{10^3 \; \rm cm^{-3}} \right)^{-1} \left(\frac{v_{\rm w}}{10^3 \; \rm km \; s^{-1}}\right)^3 \; \rm{yr}.
\end{split}
\end{equation}
\noindent 
Equating Equations~\ref{eqn:tf} and \ref{eqn:tcool} yields $v_{\rm cr}$ and is given by

\begin{equation}
\label{eqn:vcr}
\begin{split}
v_{\rm cr} &\equiv \left( \frac{\dot{M}_{\rm w} \rho_0}{4\pi C^2_1}\right)^{1/9} \\
	&= 114 \left( \frac{\dot{M}_{\rm w}}{10^{-9} \; \rm{\msun\ yr^{-1}}}\right)^{1/9} \left( \frac{n_0}{10^3 \; \rm cm^{-3}}\right)^{1/9} \; \rm km \; s^{-1}.
\end{split}
\end{equation}
\noindent
Hence, we find that B stars contained within dense, molecular clouds belong to the ``fast winds" scenario and therefore the wind bubbles that they drive are adiabatic and energy-conserving before they become pressure confined by the ambient material (e.g., see Table~\ref{tab:stars}).

\subsection{Relation between $\dot{M}_{\rm w}$ and Shell Expansion for Energy-driven Wind Feedback}
\label{sec:mdot}

As previously described, studies of the observed shells and ``bubbles" found in regions of low- and intermediate-mass star formation have only considered momentum-conserving stellar wind feedback to explain the shell dynamics \citep[e.g.,][]{Arce2011a, Feddersen2018a, Offner2015a, Offner2018a}. These studies concluded that high mass-loss rates of $\dot{M}_{\rm \star} \gtrsim 10^{-7} \rm{\msun\ \; yr^{-1}}$ are required to drive these structures, however the expected mass-loss rates for radiatively-driven winds from such stars are orders of magnitude lower. As shown in Section~\ref{sec:bubble}, the wind bubbles blown by intermediate- and high-mass stars are adiabatic and therefore the wind kinetic energy is not instantly lost via radiative cooling and instead is transferred to the ISM. Here we demonstrate that energy-driven wind feedback requires much lower mass-loss rates to drive such bubbles around young, intermediate-mass stars and are in agreement with those expected from radiatively-driven main-sequence winds.

\citet{Weaver1977a} developed the classical model for adiabatic wind bubbles driven by a central source, in which the dynamics of the swept-up material are driven by the gas pressure of the low-density, hot shock heated gas. Eventually, the swept-up ambient material experiences radiative losses and collapses to a thin shell but the shocked wind region within the shell still conserves energy. Using this classical model, we derive the relationship for the required mass-loss rate and the shell expansion dynamics for energy-conserving wind feedback before the swept-up shell experiences significant radiative losses (i.e., during the Sedov-Taylor phase). 

During the early bubble evolution the mass in the shell will be dominated by the swept-up ISM material and the integrated wind mass will contribute negligibly to the mass in the shell. Therefore, the mass contained within the shell is given by

\begin{equation}
\label{eqn:msh}
    M_{\rm sh} =\frac{4}{3} \pi \rho_0 R^3_{\rm sh}
\end{equation}
\noindent
where $\rho_0$ is the density of the average cloud density, $R_{\rm sh}$ is the shell radius, and we assume the shell is roughly spherical. \citet{Weaver1977a} find that the thermal energy of the shocked wind material contained within the adiabatic bubble that drives the shell is 

\begin{equation}
\label{eqn:eb}
E_{\rm b,\, Th} \equiv \frac{3}{2} P_{\rm b} V_{\rm b} = \frac{5}{11} \left( \frac{1}{2} \dot{M}_{\rm w} v^2_{\rm w}\right) t_{\rm w}
\end{equation}
\noindent 
where $P_{\rm b}$ and $V_{\rm b}$ are the total pressure and volume of the bubble, respectively,  $\dot{M}_{\rm w}$ is the mass-loss rate, $v_{\rm w}$ is the wind velocity, and $t_{\rm w}$ is the age of the shell.

In this scenario, the remaining wind kinetic energy is responsible for driving the shell and therefore the kinetic energy of the shell is given by
\begin{equation}
\label{eqn:esh}
E_{\rm sh, KE} \equiv \frac{1}{2}M_{\rm sh} V^2_{\rm sh} = \frac{6}{11} \left(\frac{1}{2} \dot{M}_{\rm w} v^2_{\rm w}\right) t_{\rm w}
\end{equation}
\noindent
where $V_{\rm sh}$ is the swept-up shell velocity.  The early shell expansion follows a self-similar solution similar to the Taylor-Sedov blast wave solution except now we consider a source that injects energy at a constant rate instead of an initial blast, $R_{\rm s} \propto t^{3/5}_{\rm w}$,  since radiative losses in the shell are negligible and therefore the shell expansion is in the energy-conserving phase \citep{Castor1975a, Weaver1977a}. This phase corresponds to a shell velocity of $V_{\rm sh}= 3/5 (R_{\rm sh}/t_{\rm w})$, yielding:

\begin{equation}
    M_{\rm sh} V^2_{\rm sh} = \frac{18}{55} \left(\dot{M}_{\rm w} \dot{v}^2_{\rm w}\right) \frac{R_{\rm sh}}{V_{\rm sh}}
\end{equation}
The resulting mass-loss rate during the energy conserving phase is given by
\begin{equation}
    \dot{M}_{\rm w} = \frac{110 \pi \rho_0 V_{\rm sh}^3 R_{\rm sh}^2}{27 v^2_{\rm w}}
\end{equation}
where we have replaced the mass of the shell with Equation~\ref{eqn:msh}. In terms of fiducial quantities motivated by the observed shells in star-forming regions, the mass-loss rate can be normalized as:
\begin{equation}
    \label{eqn:mdotw0} 
    \begin{split}
    \dot{M}_{\rm w} =& \; 4.52 \times 10^{-10} \left(\frac{n_0}{10^3 \; \rm{cm^{-3}}}\right) \left(\frac{V_{\rm sh}}{1 \;\rm  km \; s^{-1}}\right)^3 \\
     & \left(\frac{R_{\rm sh}}{1 \; \rm pc}\right)^2 \left(\frac{v_{\rm w}}{10^3 \;\rm  km \; s^{-1}}\right)^{-2} \; \rm{\msun \; yr^{-1}}
    \end{split}
\end{equation}
\noindent
where $n_0 =\rho_0/\mu_{\rm H}$.  We highlight that the mass-loss rate relation given by Equation~\ref{eqn:mdotw0} agrees with the mass-loss rates expected for high-mass and intermediate-mass stars (e.g., see Table~\ref{tab:stars}).

We note that \citet{Offner2015a} derive a similar relation for the momentum-conserving case in which the shell expansion rate is purely driven by the momentum in the wind:
\begin{equation}
\label{eqn:psh}
M_{\rm sh} V_{\rm sh} = (\dot{M}_{\rm w} v_{\rm w}) t_{\rm w}
\end{equation}
(we refer the reader to their Section~2.4). Comparison of Equations~\ref{eqn:esh} and \ref{eqn:psh}, show that the mass-loss rate derived for energy conserving feedback, as compared to momentum conserving feedback, is reduced by a factor of $\sim \left( V_{\rm sh}/v_{\rm w}\right)$, which is $\approx 2-3$ orders of magnitude lower than the mass-loss relation derived by \citet{Offner2015a}. 

\section{Simulation Details}
\label{sec:methods}

\begin{table*}
	\begin{center}
	\caption{
	\label{tab:stars}
Stellar Properties
}

	\begin{tabular}{ c  c  c  c  c  c  c  c c}
	\hline
	$M_{\rm \star} \, [\msun]$\tablenotemark{a} & $R_{\rm \star}\, [\rsun]$\tablenotemark{b} & $v_{\rm w}$ [$10^3$ km/s]\tablenotemark{c} &  $\vinf$-Factor\tablenotemark{c} &
	$\dot{M}_w \, [\msun \, \rm{yr}^{-1}]$\tablenotemark{c} & $L_{\rm \star} \, [10^3 \, \lsun]$\tablenotemark{b} & $\teff$ [kK]\tablenotemark{d} & 
	$v_{\rm cr}$ [km/s]\tablenotemark{e}
	\\
	\hline
	13.06 & 4.62 & 2.70 & 2.6 & $2.00 \times 10^{-9}$ & 12.7 & 28.5 &  121\\
	11.47 & 4.27 & 2.63 & 2.6 & $7.62 \times 10^{-10}$ & 8.56 & 26.9  & 109 \\
	7.08 & 3.22 & 1.19 & 1.3 & $4.04 \times 10^{-10}$ & 1.78 & 20.9 & 102 \\
	6.05 & 2.94 & 1.15 & 1.3 & $1.38 \times 10^{-10}$ & 1.04 & 19.1 &  90.1 \\
	5.19 & 2.69 &1.11 & 1.3 & $4.69 \times 10^{-11}$ & 0.61 & 17.5  & 80.0 \\
	\hline
\end{tabular}
\tablenotetext{a}{We denote stars with masses $\gtrsim 8 \; \rm \msun$ as early type B-stars (B0-B3).}
\tablenotetext{b}{ ZAMS values calculated from \citet{Tout1996a}.} 
\tablenotetext{c}{ Value calculated from \citet{Vink2001a} as described in Section~\ref{sec:wind}.} 
\tablenotetext{d}{$T_{\rm eff}$ calculated via $L_{\rm \star} =  4\pi R^2_{\rm \star} \sigma T^4_{\rm eff}$} 
\tablenotetext{e}{$v_{\rm cr}$ calculated from Equation~\ref{eqn:vcr}.}
	\end{center}
\end{table*}

In this study, we simulate how stellar feedback from fast, isotropic main-sequence winds from young intermediate- and high-mass stars (i.e., from $\sim5-13 \; M_{\rm \odot})$ may drive pc-scale bubbles in molecular clouds using the \orion\ adaptive mesh refinement (AMR) constrained-transport gravito-magneto-hydrodynamics (MHD) simulation code \citep{Li2012a}, which is based on the finite volume scheme implemented in \pluto\ \citep{Mignone2012a}. \orion\ also contains Lagrangian sink particles that represent stars \citep{Krumholz2004a} coupled to a sub-grid prescription that models feedback from isotropic stellar winds \citep{Offner2015a}. We summarize the numerical methods and our AMR refinement criteria in Section~\ref{sec:numeth}. In Section~\ref{sec:ics}, we present the initial and boundary conditions for our simulations and describe our stellar wind feedback sub-grid model in \ref{sec:wind}.

\subsection{Numerical Methods and Refinement Criteria}
\label{sec:numeth}
We perform two simulations that are identical in every way except that the first simulation, which we refer to as \isoRun, uses an isothermal equation of state, which assumes the energy injected by winds into the surrounding ISM is instantly radiated away. In this scenario, the wind bubble dynamics are driven only by momentum feedback. The second simulation, \adiaRun, includes an adiabatic equation of state and realistic gas and dust cooling to more accurately model the wind thermodynamics and follow the adiabatic expansion of the hot bubbles produced by the shock-heating (i.e., thermalization) of the injected stellar winds.  

For each simulation we treat the gas as an ideal gas so that the gas pressure is given by
\begin{equation}
P=\frac{\rho k_{\rm B}T}{\mu m_{\rm H}} = \left( \gamma-1\right) \rho e_{\rm T},
\end{equation}
where $\rho$ is the gas density, $T$ is the gas temperature, $\mu=2.33$ is the mean molecular weight appropriate for molecular hydrogen at solar composition,  $\gamma$ is the ratio of specific heats, and $e_{\rm T}$ is the thermal energy of the gas per unit mass. We adopt $\gamma = 1.0001$ for \isoRun, appropriate for an isothermal equation of state, and adopt  $\gamma=5/3$ for \adiaRun, which is appropriate for molecular gas at temperatures too low to excite the rotational levels of H$_2$ \citep[e.g.,][]{Rosen2016a} and the hot shock-heated gas produced by stellar wind feedback \citep{Weaver1977a}. 

For \isoRun\ our chosen value of $\gamma$ corresponds to efficient cooling, such that the wind's thermal and kinetic energies contributes little to the expansion.
To model radiative cooling for \adiaRun\ we supply a cooling function, $\Lambda (T)$, to take into account radiative losses for gas ranging from 10~K$-10^9$~K. For gas with $T < 10^4$~K, $\Lambda (T)$ is given by the analytical cooling function from \citet{Koyama2002} (their Equation~4). At higher temperatures we use the cooling function calculated by \chianti\ where we have assumed the gas is at solar metallicity and in collisional ionization equilibrium \citep{Dere1997a}.

For each simulation we begin with a base grid with volume (5 pc)$^3$ discretized by $256^3$ cells and allow for two levels of refinement, resulting in a maximum resolution of $10^3$ au. As the simulation evolves, the AMR algorithm automatically adds and removes finer grids based on certain refinement criteria set by the user. We refine cells if they meet at least one of the following criteria: (1) any cell where the density in the cell exceeds the Jeans density given by

\begin{equation}
\label{eqn:rhoj}
\rho_{\rm max,J} = \frac{\pi J^2_{\rm max} c_{\rm s}^2}{G \Delta x^2_l} \left( 1 + \frac{0.74}{\beta^2}\right),
\end{equation}
\noindent
where $c_{s}=\sqrt{kT/\mu m_{\rm{p}}}$ is the thermal sound speed, $\Delta x_l$ is the cell size on level $l$, $\beta=8\pi \rho c_s^2/B^2$ is the plasma parameter (i.e., the ratio of the thermal gas pressure to the magnetic pressure), and $J_{\rm max}$ is the maximum allowed number of Jeans lengths per cell, which we set to 1/8, following the MHD Truelove Criterion \citep{Myers2013a};  (2) any cell that is located within at least eight cells of a star particle, which ensures the star particle is refined to the highest level; and (3) any cell within which the density gradient exceeds $\nabla \rho > 10 \rho/\Delta x_{l}$ to ensure the dense shells are resolved.

\subsection{Initial and Boundary Conditions}
\label{sec:ics}
Following \citet{Offner2015a}, the simulations presented in this work are modeled as a periodic box of length side $L=5$~pc that represent a patch of a molecular cloud with total mass $3767 \, \msun$, corresponding to an average density of $\rho_0 = 2.04 \times 10^{-21} \, \rm{g \, cm^{-3}}$. We set the initial gas temperature to 10~K and the initial cloud Mach number to $\mathcal{M}_{\rm 3D}=10.5$ (with corresponding 1D velocity dispersion $\sigma_{\rm 1D}=1.14 \, {\rm km \,s^{-1}}$), following the line-width size relation \citep[e.g.,][]{McKee2007a}. 

We include turbulence by driving the initial gas velocities ($v_x$, $v_y$, and $v_z$) with a velocity power spectrum, $P(k) \propto k^{-2}$, as is expected for supersonic turbulence \citep{Padoan1999a, Boldyrev2002a, Cho2003a, Kowal2007a}, for three crossing times, $t_{\rm cr} = L/\sigma_{\rm 1D}$, so that it reaches a steady state \citep{Federrath2013a}. We include modes between $k_{\rm min} = 1$ to $k_{\rm max} = 2$, corresponding to a linear size of $L/2-L$ and take the turbulence mixture of gas to be 1/3 compressive and 2/3 solenoidal, which is consistent with the natural mixture of turbulent modes for a 3D fluid \citep{Kowal2007a, Kowal2010a}. The turbulent cascade populates motions on smaller scales, producing the expected power spectrum for supersonic turbulence. During the driving phase we set $\gamma=1.0001$ for both simulations so that this phase is purely isothermal. When gravity is turned on we keep $\gamma$ set to this value for run \isoRun\ but change $\gamma$ to 5/3 for run \adiaRun\ and include the cooling function, $\Lambda(T)$, described above. 

The simulations start with a uniform magnetic field,  $\mathbf{B} =  B_{0} \hat{z}$,  which becomes disordered over the turbulent driving period. We set the initial $\beta =0.1$, such that the initial field strength is $B_0 =13.4 \, \rm{\mu G}$, which is a typical average value for molecular clouds \citep{Crutcher2012a}. After the turbulence driving phase we randomly place five stars with masses ranging from $5.19-13.06 \, \msun$ within the computational domain and turn on gravity and the launching of stellar winds for each star, which we describe next. We list their stellar properties in Table~\ref{tab:stars}. 
We run both simulations for 0.2 Myr once feedback from winds is turned on.

\begin{figure*}
\centerline{\includegraphics[trim=2.5cm 1.25cm 1.0cm 1.25cm,clip,width=0.95\textwidth]{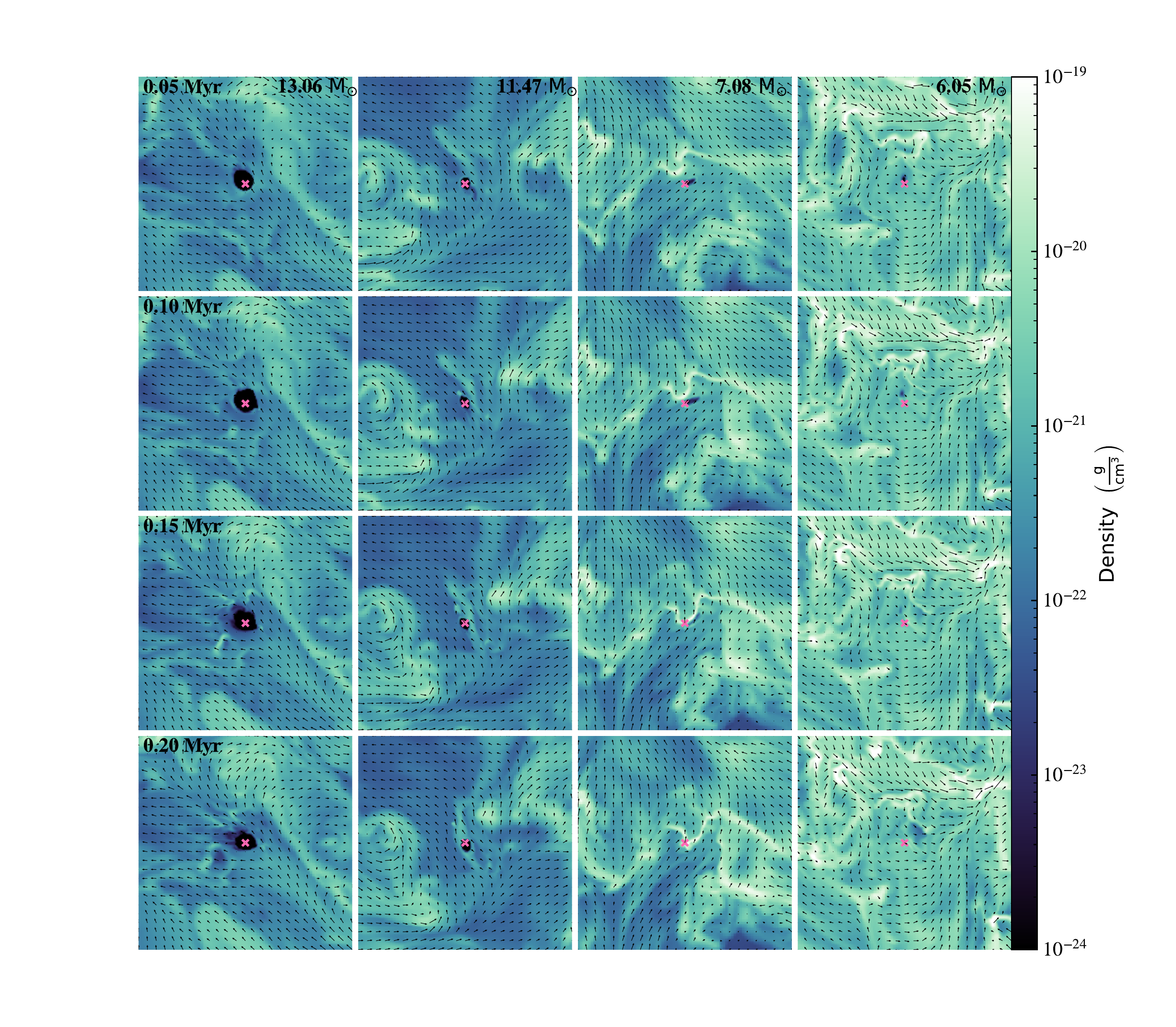}}
\caption{
\label{fig:isoBubbles}
Time evolution of the density slices along the $yz-$plane for run \isoRun\ with magnetic field vectors over-plotted. Each column shows the time evolution for the four most massive stars. The time of the simulation and mass of each star are given in the top-left of each row and top-right of each column, respectively. Each panel is $(2$ pc$)^2$, and the pink X at the center of each panel denotes the location of the star. Momentum-driven wind feedback from the high-mass stars drives sub-pc bubbles but has little effect for intermediate-mass stars. Eventually, the bubbles become pressure confined, and the expansion stalls within $\sim$0.05-0.1 Myr for the two most massive stars.
}
\end{figure*}

\subsection{Stellar Wind Model}
\label{sec:wind}
We use the sub-grid stellar wind feedback prescription described in \citet{Offner2015a}, which injects the winds isotropically within the 8 nearest zones to the star. Here the total mass, momenta, and kinetic and thermal energies injected within this region equate to the amount injected by the stellar wind from each star particle. Following \citet{Offner2015a} we set the wind gas temperature to $10^4$~K, appropriate for a fully ionized gas. 

For the stellar wind mass-loss rates we adopt the \citet{Vink2001a} mass-loss recipe. This formalism, which depends on the stellar properties, is adapted from Monte Carlo simulations that follow the fate of a large number of photons from below the stellar photosphere and calculate the radiative acceleration  (i.e., launching) of the wind material for stars with effective temperature $T_{\rm eff} \ge$ 12,500~K. \citet{Vink2001a} show that the wind mass-loss rates experience a jump around $T_{\rm eff} \approx$ 25,000~K, known as the bi-stability jump. This effect is due to a change in the ionization state in the lower stellar photosphere (i.e., Fe iv recombines to form Fe iii) leading to Fe ions that are more efficient line drivers on the hot-side of the bi-stability jump. \citet{Vink2001a} finds that on the hot-side of the bi-stability jump the mass-loss rate increases by a factor of $\sim5$. 

Following Equation~15 from \citet{Vink2001a} we first compute the bi-stability jump temperature to compute the correct stellar wind mass-loss rate defined next. For the cool side of the bi-stability jump the mass-loss rate is

\begin{equation}
\label{eqn:mdot1}
\begin{split}
\log_{10}(\dot{M}_{\rm \star}) = &-6.688 + 2.210 \log_{10} \left(\frac{L_{\rm \star}}{10^5 \, \lsun} \right)\\
					       & - 1.339 \log_{10}\left( \frac{M_{\rm \star}}{30 \, \msun}\right) \\
					      &  - 1.601 \log_{10} \left( \frac{v_{\rm \infty}/v_{\rm esc}}{2.0} \right) \\
				&+ 1.07 \log_{10} \left( \frac{T_{\rm eff}}{20 \, \rm kK}\right)
\end{split}
\end{equation}
\noindent
where $M_{\rm \star}$ is the stellar mass, $R_{\rm \star}$ is the stellar radius, $v_{\rm esc}=\sqrt{2GM_{\rm \star}/R_{\rm \star}}$ is the escape speed of the star, and $\vinf$ is the wind velocity. For stars with temperatures above the hot side of the bi-stability jump (i.e., $\teff \gtrsim 25 \; \rm K$) the mass-loss rate is given by 
\begin{equation}
\label{eqn:mdot2}
\begin{split}
\log_{10}(\dot{M}_{\rm \star}) = &-6.697 + 2.194 \log_{10} \left(\frac{L_{\rm \star}}{10^5 \, \lsun} \right)\\
					       & - 1.313 \log_{10}\left( \frac{M_{\rm \star}}{30 \, \msun}\right) \\
					        &- 1.226 \log_{10} \left( \frac{v_{\rm \infty}/v_{\rm esc}}{2.0} \right) \\
				&+ 0.933 \log_{10} \left( \frac{T_{\rm eff}}{40 \, \rm kK}\right) \\
				& - 10.92 \left( \log_{10} \left( \frac{T_{\rm eff}}{40 \, \rm kK}\right)  \right)^2.
\end{split}
\end{equation}
\noindent
Both of these mass-loss formulas assume solar metallicity. We assume the wind velocities, $\vinf$, are $1.3 v_{\rm esc}$ and $2.6 v_{\rm esc}$ for stars on the cool-side and hot-side of the bi-stability jump, respectively, following values of $\vinf/v_{\rm esc}$ determined by both theory and observations of winds from B and O stars \citep[][and references therein]{Vink2001a}. We assume the stars modeled in our work are zero age main sequence (ZAMS) stars and therefore use the ZAMS luminosity and radius calculated with the formulae from \citet{Tout1996a} to compute $T_{\rm eff}$. We use these values for the mass-loss rates given by Equations~\ref{eqn:mdot1} and \ref{eqn:mdot2}.

\section{Results}
\label{sec:results}

\subsection{Bubble Structure and Evolution}
\label{sec:bubbles}

\begin{figure*}
\centerline{\includegraphics[trim=2.5cm 1.25cm 1.0cm 1.25cm,clip,width=0.95\textwidth]{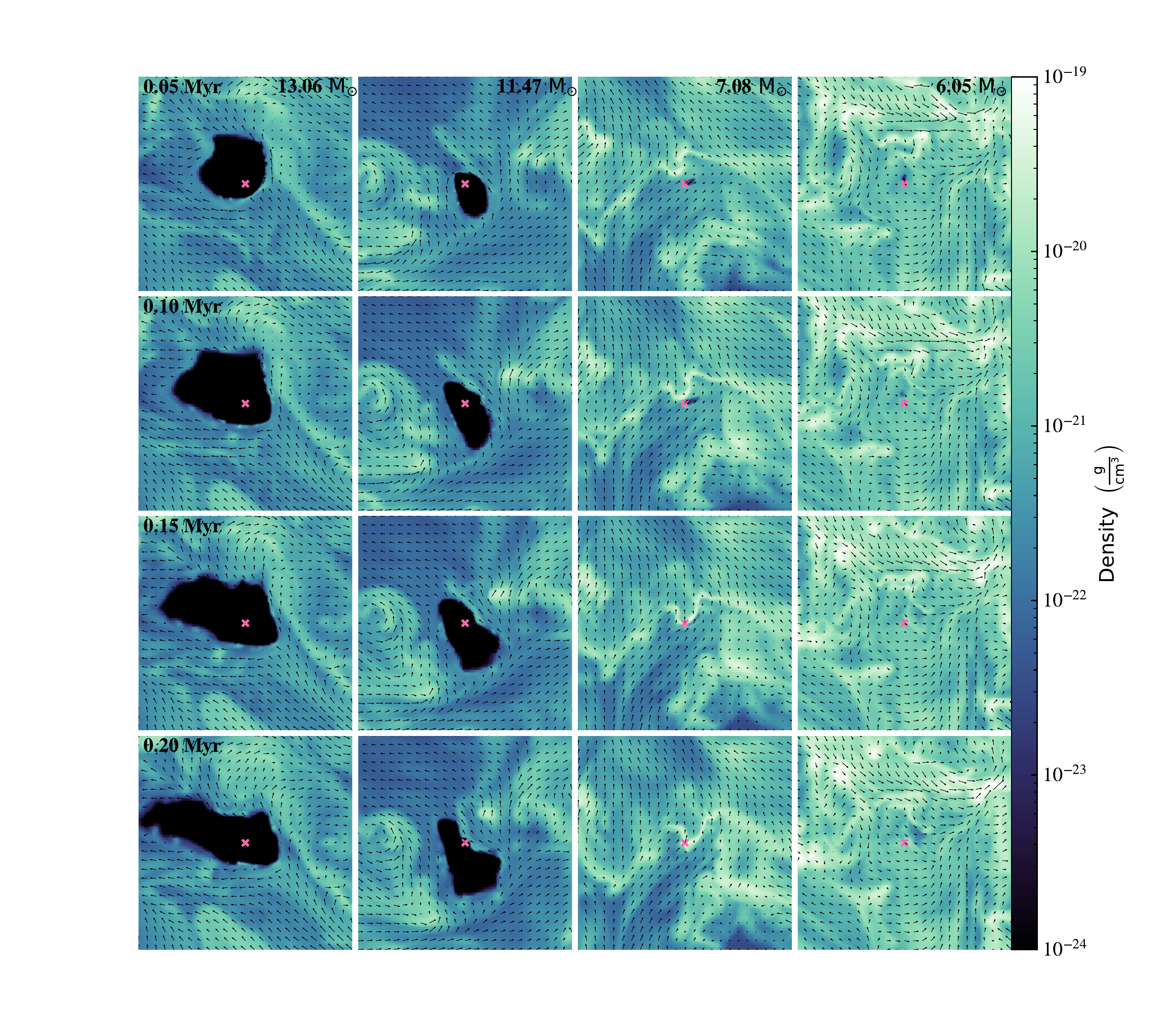}}
\caption{
\label{fig:adiaBubbles}
Same as Figure~\ref{fig:isoBubbles} except for \adiaRun. This figure demonstrates that energy-driven wind feedback from high-mass stars drives larger wind-driven bubbles than momentum-driven wind bubbles. However, for intermediate-mass stars the kinetic energy injection by winds has little effect on driving wind driven bubbles and is therefore likely lost via radiative cooling.
}
\end{figure*}

Figures~\ref{fig:isoBubbles} and \ref{fig:adiaBubbles} show the time evolution of the expanding shells driven by stellar wind feedback (i.e., wind-driven bubbles) for stars with mass $>6 \, \rm \msun$ for \isoRun\ and \adiaRun, respectively. Each column shows the time evolution (from top to bottom) of the bubble structure for each star, with the star centered in each panel. 
Comparison of the columns in Figure~\ref{fig:isoBubbles} for \isoRun\ shows that wind momentum feedback for intermediate-mass stars with masses $\lesssim 7.08 \, \msun$ (i.e., late B stars) drives a small non-spherical structure at early times. However, these structures are short-lived because they  become confined at late times due to gravity and the thermal, magnetic, and turbulent pressure of the ambient cloud. 
Therefore, we find that momentum-driven expansion produced by intermediate-mass (late-type B) stellar winds is not sufficient to drive substantial bubbles in molecular clouds. 

The two left-most columns of Figure~\ref{fig:isoBubbles} show the bubble evolution for the two (massive) early-type B stars modeled, both of which lie on the hot side of the bi-stability jump (see Table~\ref{tab:stars}). Small sub-parsec (i.e., $\sim0.05-0.1$ pc) spherical bubbles appear around these stars, suggesting that the higher mass-loss rates and wind velocities (i.e., by a factor of $\sim 4$ as compared to the intermediate-mass stars modeled) are able to drive bubbles around early type B-stars. We note that these bubble radii for the high-mass stars in our sample agree with the small sub-parsec bubbles observed around low- to intermediate mass YSOs but are still smaller than the majority of bubbles identified in Perseus, Taurus, and Orion \citep[e.g.,][]{Arce2011a, Lili2015a, Feddersen2018a, Xu2020a}. Additionally, the bubbles become pressure confined, and the expansion stalls within $\sim0.05-0.1$ Myr. These bubbles remain roughly spherical for the duration of the simulation. 

\begin{figure}
\centerline{\includegraphics[trim=0.2cm 0.2cm 0.2cm 0.2cm,clip,width=0.95\columnwidth]{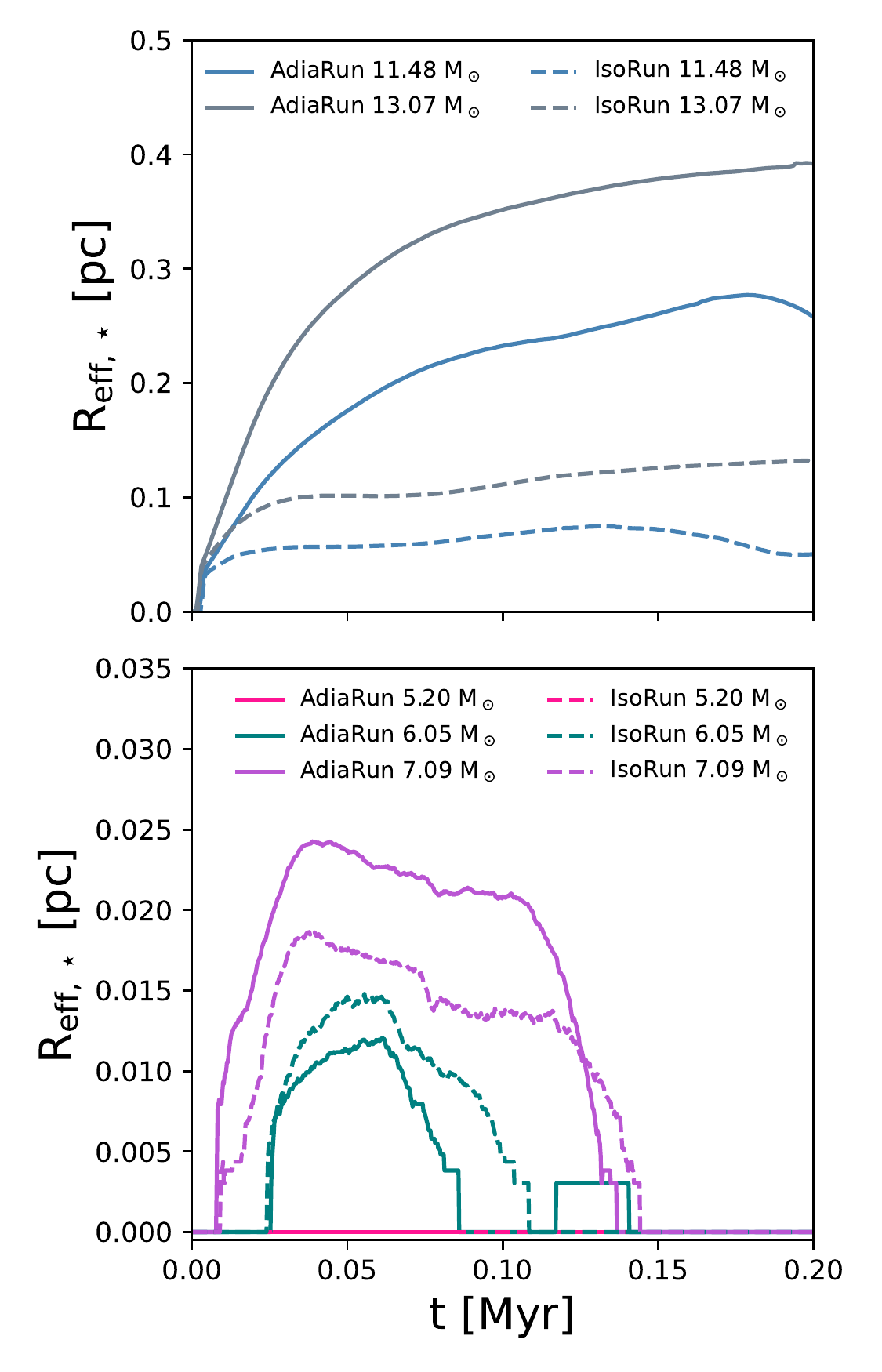}}
\caption{
\label{fig:Reffstars}
Effective radii for the bubbles driven by wind feedback for the massive stars (top panel; $M_{\rm \star} > 8 M_{\rm \star}$ and intermediate mass stars (bottom panel; $M_{\rm \star} < 8 M_{\rm \star}$) for runs \isoRun\ (dashed lines) and \adiaRun\ (solid lines). The bubbles that surround the high-mass stars become pressure-confined for \isoRun\ within 0.05-0.1 Myr and the bubbles sizes for the intermediate mass stars in both simulations are roughly similar demonstrating that energy-driven wind feedback cannot produce the observed shells.}
\end{figure}

Next, we consider how the kinetic wind energy ($\dot{E}_{\rm w} = \frac{1}{2} \dot{M}_w v^2_w$) is transferred to the ISM via shock heating of the fast, isotropic stellar winds, which generates hot, low density gas that undergoes little radiative losses and therefore cools via adiabatic expansion, as shown in Figure~\ref{fig:adiaBubbles} for run \adiaRun. Given that the wind velocities are high, the wind-blown bubbles simulated here are in the ``fast wind" regime and are adiabatic once the swept-up gas dominates the gas injected by stellar winds (e.g., see Section~\ref{sec:bubble}). Eventually these bubbles will stop expanding once the pressure of the ambient medium becomes significant and they become pressure confined \citep{Koo1992a}. The two left columns of  Figure~\ref{fig:adiaBubbles} show that for the two massive (early-type B) stars simulated inclusion of energy feedback from stellar winds, rather than assuming it is lost instantly via radiation, drives larger scale bubbles around these stars, with sizes up to $\sim1$ pc. Therefore, we find that the energy-driven phase accounts for most of the bubble expansion. These bubble sizes are in better agreement with those observed in Perseus, Taurus, and Orion where the typical bubble sizes range up to $\sim$1 pc. However, the structures that form around the intermediate-mass stars for \adiaRun\ (next three columns of Figure~\ref{fig:adiaBubbles}) look nearly identical to those formed in \isoRun. Therefore, we conclude that the mass-loss rates and velocities characteristic of intermediate-mass stars do not efficiently drive bubbles in either expansion limit. Hence, the observed shells must either be produced by winds from more massive stars or by another physical process (see discussion in Section~\ref{sec:sources}). 

Inspection of the shells produced by wind feedback from the two most massive stars in Figure~\ref{fig:adiaBubbles} show that the shells are not entirely smooth and have density enhancements or ``knots" at the interface between the dense shell and low-density bubble. These density enhancements in the shells can form via Kelvin-Helmholtz instabilities that occur as the hot gas flows into the cold surrounding interstellar material \citep{McKee1984a, Strickland1998a, Nakamura2006a, GallegosGarcia2020a} and/or via Rayleigh-Taylor instabilities that occur at the interface between two fluids of different densities in which the lighter fluid (e.g., bubble interior) pushes into the heavier fluid \citep[e.g., bubble shell;][]{Rosen2016a}. Both types of instabilities can eventually grow non-linearly and efficiently mix the hot and cold gas \citep{Rosen2014a, Lancaster2021a}.  However, at least for the time simulated here, the growth of these instabilities and subsequent mixing are suppressed, and this effect is likely due to magnetic tension in the shell \citep[e.g., see review by][]{Krumholz2019b}. As the bubbles expand, the magnetic field lines become parallel to the interface between the hot bubble and dense shell. Magnetic field lines parallel to the hot-cold interface will strongly delay or suppress mixing via Kelvin-Helmholtz and Rayleigh-Taylor instabilities \citep{Stone2007a, Arthur2011a, McCourt2015a, Banda2018a}. \citet{Offner2015a} saw similar structures at the interfaces of their momentum-conserving wind bubbles and suggest that these knotted features are likely due to the magnetic kink instability. Aside from this instability, we see negligible mixing at the hot-cold interface of the wind driven bubbles produced by energy-driven wind feedback.

\begin{figure*}
\centerline{\includegraphics[trim=0.2cm 0.2cm 0.2cm 0.2cm,clip,width=0.95\textwidth]{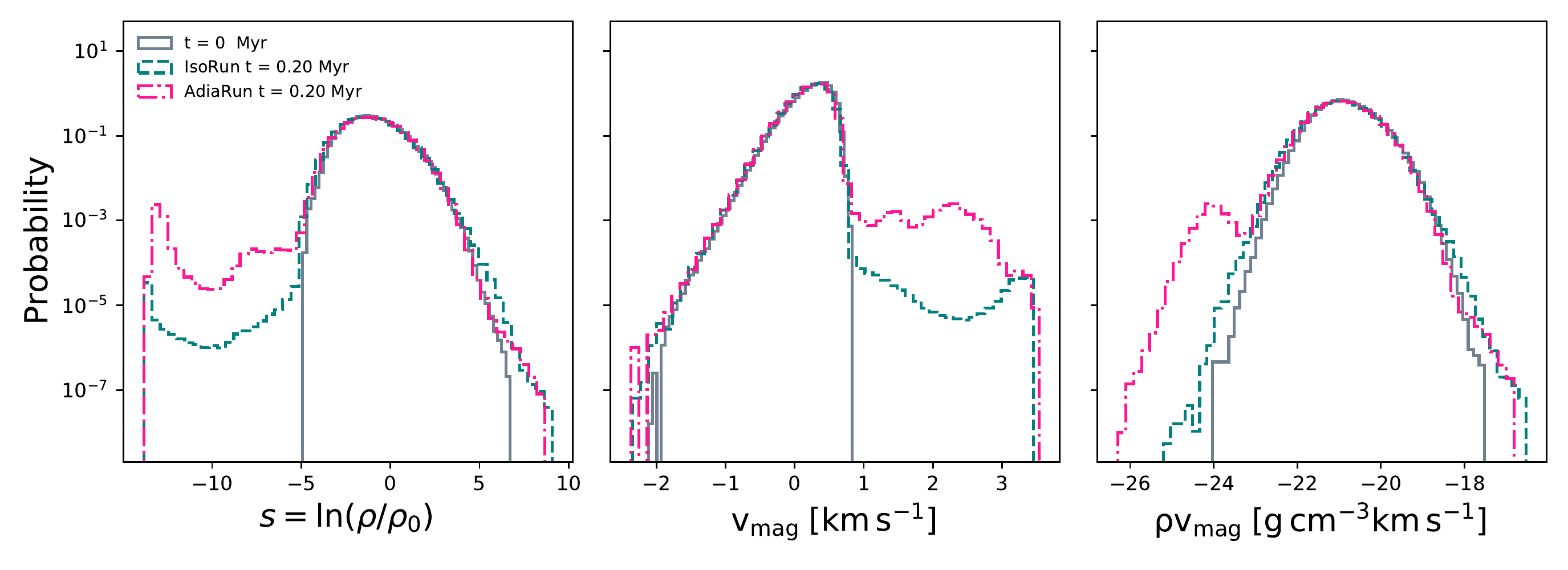}}
\caption{
\label{fig:hist}
Histograms of the gas density (left), velocity (middle), and momentum (right) for the initial snapshot (gray lines) and final snapshots at $t=0.2$ Myr for runs \isoRun\ (teal lines) and \adiaRun\ (pink lines). We parameterize the density as $s=\ln(\rho/\rho_0)$ where $\rho_0$ is the initial cloud density since ISM turbulence follows a log-normal distribution. These panels demonstrate that wind feedback drives high velocity, low density gas within the bubbles and that this effect is enhanced for adiabatic energy-driven wind bubbles.}
\end{figure*}

Figure~\ref{fig:Reffstars} shows the effective bubble radii for all stars in both simulations. The top panel shows the individual bubble radii that surround the high-mass (early-type B) stars and the bottom panel shows the bubble radii for the intermediate-mass (late-type B) stars. We compute the radius for each bubble by calculating $R_{\rm eff, \star} = \left[\Sigma_i \delta x_{T>1000 K}^3/(4\pi/3)\right]^{1/3}$ in a 1.1 pc region surrounding each star, where we sum over all cells of width $\delta x$ that contains gas with $T>10^3$ K. This figure shows that the two most massive stars in these simulations produce substantial bubbles, whereas the three intermediate-mass stars produce bubble sizes (except that of the 5.20 $\msun$ star) that are approximately a few to ten times smaller than those driven by the more massive stars regardless of whether the expansion is isothermal or adiabatic. However, the bubble radii of the intermediate-mass stars for both adiabatic and isothermal runs are comparable. 

We conclude that the adiabatic expansion of bubbles driven by main sequence stellar winds from high-mass stars is important, while winds from intermediate-mass stars have only a minor impact. Likewise, for the intermediate mass stars modeled here, the bubbles expand at early times but then shrink and eventually disappear at late times due to gravity and external pressure.  We also note that the 5.2 $\msun$ star modeled here, which has a mass-loss rate $\sim2.5$ times lower than the $6.05~M_\odot$ star, does not drive a bubble regardless of the equation of state modeled. 

\subsection{Gas Properties}
Stellar feedback alters the mass density and dynamics of gas in star-forming environments \citep[e.g.,][]{Lopez2014a, Menon2021a, GallegosGarcia2020a, Olivier2021a}. Figure~\ref{fig:hist} shows normalized histograms of the gas density (left panel), velocity magnitude (middle panel), and momentum (right panel) for \isoRun\ (teal dashed lines) and \adiaRun\ (pink dot-dashed lines) for the final snapshots at $t=0.2$ Myr with the initial histograms over-plotted (gray solid lines). Comparison of each final histogram for both simulations with the initial snapshot shows that the peak and width of the distribution of these quantities show little change. However, the tails of the gas density, velocity, and momentum distributions are dramatically different as the simulation evolves because stellar winds produce large regions with high-velocity, low-density gas. 

ISM turbulence follows a log-normal distribution \citep[e.g., ][]{McKee2007a}.
Inspection of the gas density distribution in Figure~\ref{fig:hist} shows that gas becomes denser at the end of both simulations due to the gravitational contraction of dense filaments and swept-up material in the wind-driven shells.
Wind feedback produces low-density gas within these bubbles; however this is more prominent in \adiaRun\ because the bubbles are larger.  
The right-most panel shows a peak for low momentum gas for \adiaRun\ due to the fast but low-density, shock-heated gas contained within the wind-driven bubbles, which is not seen in \isoRun.

\subsection{Driven Turbulence}
Feedback from stellar winds may also sustain turbulence in molecular clouds \citep[e.g.,][]{Arce2011a, Offner2015a, Lili2015a, Feddersen2018a, GallegosGarcia2020a,Xu2020a}. As described in Section~\ref{sec:ics}, we no longer drive turbulence at the beginning of the simulations (i.e., when we turn on gravity and stellar winds) and therefore allow turbulence to decay as the simulation evolves. In the absence of stellar feedback or other driving sources, turbulence will decay within $ t_{\rm D} \sim L/\sigma_{\rm 1D}$ \citep{Goldreich1995a}. We note that for our simulation parameters we have $t_{\rm D} \approx 4.2$ Myr, which is much longer than the timescale that we evolve with winds, and therefore the initial turbulence does not significantly decay. 

\begin{figure*}
\centerline{\includegraphics[trim=0.2cm 0.2cm 0.2cm 0.2cm,clip,width=0.95\textwidth]{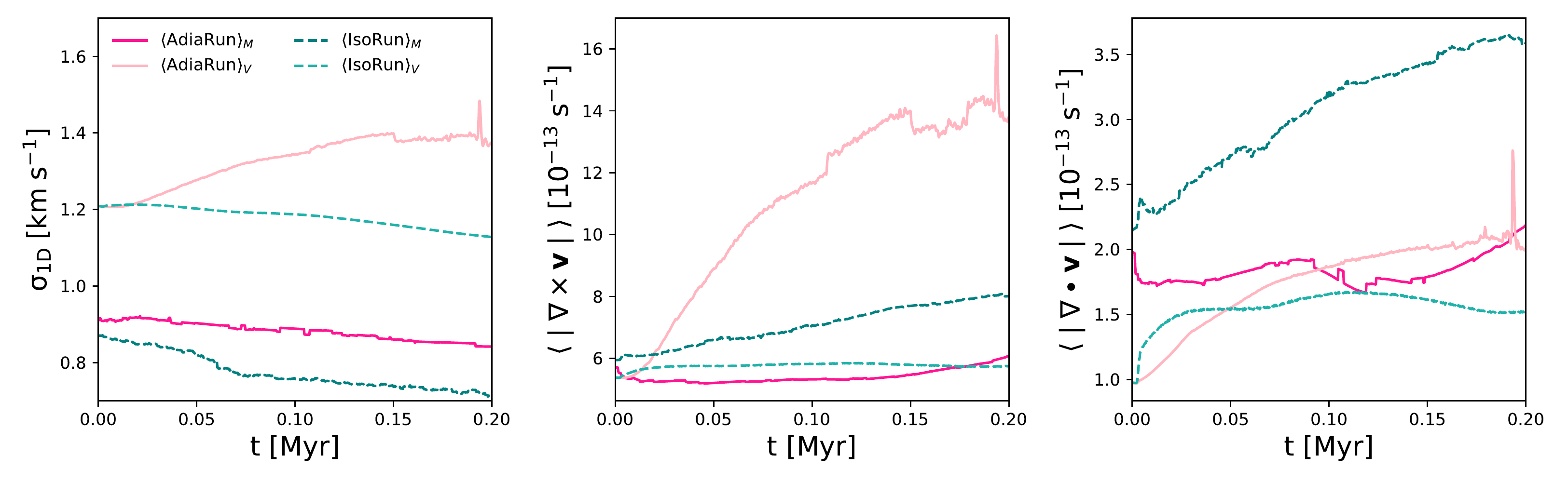}}
\caption{
\label{fig:velocity}
 Evolution of the cloud velocity dispersion (left panel), mean vorticity magnitude (middle panel), and mean velocity divergence magnitude (left panel) for runs \adiaRun\ (pink solid lines) and \isoRun\ (teal dashed lines). The darker (lighter) colors show the mass-weighted (volume-weighted) quantities. This figure demonstrates that energy-driven wind feedback is more efficient than momentum-driven wind feedback at driving solenoidal turbulence in the low-density gas carved out by stellar winds and sustaining turbulence in molecular clouds.
 }
\end{figure*}

As previously described, stellar wind feedback drives dense shells and small-scale motions, thereby offsetting the turbulent dissipation rate.  Figure~\ref{fig:velocity} shows the time evolution of the cloud mass- and volume-weighted 1D gas velocity dispersion (left panel), which illustrates how wind feedback drives turbulence within molecular clouds. The mass-weighted quantities preferentially emphasize the dense regions, including the bubble shells and dense shocks and filaments, whereas the volume-weighted quantities highlight  the fast, low-density gas inside the stellar wind-driven shells. For both simulations we see that the overall mass-weighted velocity dispersion decays slightly but that the decay is less prominent in \adiaRun. Likewise, the volume-weighted velocity dispersion declines in \isoRun\ but increases in \adiaRun\ due to the larger impact of the adiabatic expansion of the hot gas produced by the shock-heating of stellar winds. Even though the bubbles themselves have a small volume filling fraction within the simulation domain, the fast-flowing gas within the bubble interiors enhances the overall volume-weighted velocity dispersion in \adiaRun.

Turbulent driving can be described as a mix of solenoidal/stirring (i.e., non-zero vorticity - $\nabla \times \mathbf{v} \neq 0$) and compressive  (i.e., non-zero velocity divergence - $\nabla \bullet \mathbf{v} \neq 0$) motions \citep[e.g.,][]{Federrath2013a}. In Figure~\ref{fig:velocity} we show the time evolution of the cloud mass- and volume-weighted vorticity magnitude ($\mid \nabla \times \mathbf{v} \mid$; center panel) and the velocity divergence magnitude ($\mid \nabla \bullet \mathbf{v} \mid$; right panel) in Figure~\ref{fig:velocity}. In \adiaRun\ stellar winds increase the volume-weighted gas vorticity by a factor of a $\sim$few within 0.2 Myr, whereas this quantity is roughly constant throughout the simulation in \isoRun. The larger increase of the volume-weighted vorticity in \adiaRun\ is because the energy-driven bubbles are much larger than the momentum-conserving bubbles produced in \isoRun. Therefore, adiabatic expansion drives solenoidal turbulence in the shock-heated, low density gas contained within the wind-driven bubbles. In contrast, the mass-weighted vorticity magnitude, which traces the dense gas, increases only moderately.  

The right panel of Figure~\ref{fig:velocity} shows that the volume-weighted velocity divergence magnitude for \adiaRun\ increases by a factor of $\sim$2 by the end of the simulation. However, this quantity for \isoRun\ only increases by $\sim$50\% on a short time scale and remains roughly constant.  This effect is likely due to the quick confinement of the wind bubbles as shown in Figures~\ref{fig:isoBubbles} and \ref{fig:Reffstars}. The mass-weighted velocity divergence magnitude remains roughly constant for \adiaRun\ throughout the simulation but increases by a factor of $\sim$1.5 for IsoRun by the end of the simulation. This is due to the gravitational contraction of dense gas that occurs in isothermal turbulence that is no longer being actively driven by outside mechanisms. 

\section{Discussion}
\label{sec:discussion}
The purpose of this work is to determine if the momentum and kinetic energy injected by radiatively-driven winds from young, intermediate- and high-mass stars can drive $\sim$pc scale bubbles in star-forming environments like those found in Perseus, Taurus, and Orion \citep{Arce2011a, Lili2015a, Feddersen2018a}. Most notably, we find that energy-driven feedback from stellar winds, in which the kinetic energy is thermalized and generates very hot, low-density gas that then cools via adiabatic expansion rather than radiation, can drive $\sim$pc scale bubbles around high-mass stars. However, intermediate-mass stars with weaker winds do not achieve the same effect. Therefore, we find that energy-driven feedback from radiatively-driven winds produced by young, intermediate-mass stars can not explain the observed bubbles. Additionally, we find that energy-driven stellar wind feedback produces fast flowing, low-density gas that drives small-scale turbulence in molecular clouds. In the following section, Section~\ref{sec:turb}, we discuss how wind feedback drives local turbulence in molecular clouds. We then discuss the implications of energy- versus momentum-driven feedback from stellar winds in Section~\ref{sec:feedback}. Finally, we explore other possible feedback mechanisms that may be responsible for producing the shells observed around young, intermediate-mass stars in Section~\ref{sec:sources}.  

\subsection{Wind Feedback Driven Turbulence in Molecular Clouds}
\label{sec:turb}
As demonstrated throughout this work, feedback from radiatively-driven isotropic stellar winds alters the density distribution and dynamics of molecular clouds regardless of whether we consider momentum-driven (\isoRun) or energy-driven (\adiaRun) feedback. By modeling both cases, we find that energy-driven wind feedback, in which the kinetic energy from fast winds produced by massive (early-type B) stars drives $\sim$pc scale bubbles, generates solenoidal turbulence in the low-density gas carved out by stellar winds. We also find that these bubbles are larger than the momentum-conserving case, which assumes the kinetic energy from winds is instantly lost via radiation and therefore not transferred to the ISM. Additionally, the expanding bubble shells sweep up molecular gas, thereby driving compressive turbulence in the ISM. 

Previous numerical work has also studied how wind feedback can generate turbulence in molecular clouds. \citet{Offner2015a} and \citet{Offner2018a} performed similar simulations to \isoRun\  presented here but modeled wind feedback assuming much larger wind mass-loss rates ($\dot{M}_{\rm w} \sim 10^{-7} \; \rm \msun\ yr^{-1}$) and slower stellar winds ($v_{\rm w} = 200$ km/s). They found that winds do not produce clear features in turbulent statistics such as the Fourier spectra of density and momentum but do impact the Fourier velocity spectrum. \citet{Offner2018a} found that these bubbles  can transfer energy and therefore drive turbulence to larger scales within molecular clouds via magnetic waves excited by the expanding wind shells. They conclude that stellar feedback, interacting in a magnetized medium, can partially offset the global turbulence dissipation.  Likewise, we find that energy-driven feedback from weaker stellar winds increases the injected energy within the cloud and therefore is more effective than momentum-driven feedback for sustaining local turbulence in molecular clouds.

In the context of star clusters containing intermediate- and high-mass stars, \citet{GallegosGarcia2020a} modeled how the collective stellar wind bubble produced by a star cluster that contains intermediate- and high-mass stars can drive local subsonic Kolmogorov turbulence in an initially uniform (i.e., non-turbulent) medium. This turbulence is generated via mixing between the bubbles driven by winds from individual stars and subsequent mixing at the low-density, hot cluster bubble interface with the surrounding swept-up cold ISM interface.  Likewise, we find that even if the wind bubbles from nearby stars do not overlap stellar winds can still drive local turbulence within molecular clouds. In the simulations presented here, we find that stellar wind feedback preferentially drives low-density solenoidal turbulence within the individual bubbles carved out by stellar winds. 


Our results conclude that stellar wind feedback can drive turbulence locally within molecular clouds. Previous work has shown that large scale (i.e., global) turbulence is driven via gravitational torques induced by the mass flows within galaxies, shearing motions of GMCs within galaxies, and cloud-cloud collisions \citep{Tasker2009a, Goldbaum2015a}. Additionally, on individual cloud and star-forming clump scales, gravitational contraction can also drive turbulent motions within them \citep{Traficante2018a, Rosen2019a}. Therefore, the simulations presented here suggest that stellar wind feedback only drives turbulence locally and therefore these aforementioned processes are necessary for driving larger scale non-thermal motions that can help sustain turbulence in molecular clouds. 

\subsection{Energy-conserving vs. Momentum-conserving Wind Bubbles}
\label{sec:feedback}

Previous work has explored how the stellar wind energy is transferred to the ISM \citep[e.g.,][]{Weaver1977a, Harper-Clark2009a, Lopez2011a, Rogers2013a, Rosen2014a, Fierlinger2016a, GallegosGarcia2020a, Lancaster2021a}. Much of this work has been motivated by the unexpectedly weak observed X-ray emission from star clusters that host many massive stars. In this scenario, the shock-heated gas produced by wind feedback primarily radiates via X-rays due to its high temperature ($T\gtrsim 10^6$ K) and the thermal energy retained in the hot gas, as probed by the X-ray emission, is weaker than that predicted from the \citet{Weaver1977a} model (e.g., see Equation~\ref{eqn:eb}). These observations suggest that the majority of the injected wind kinetic energy is lost as the hot wind bubble, which encompasses the star cluster, expands into the surrounding cold and turbulent ISM. 

As first described by \citet{Rosen2014a} and modeled by \citet{Lancaster2021a} the expansion of the hot gas produced by stellar wind feedback in a turbulent medium can lead to instabilities that mix the cold, molecular gas into the hot bubble interior leading to mixed gas temperatures of $\sim 10^4-10^5$ K. This gas lies at the peak of the cooling curve and will quickly suffer radiative losses \citep[e.g., see Figure~1 of ][]{Rosen2014a}. This process of ``turbulent conduction" depends on how well the hot gas retained within the bubble can efficiently mix with the surrounding cold, turbulent ISM. \citet{Lancaster2021a} showed that, in the absence of magnetic fields, the interaction between the hot gas produced by the shock heating of stellar winds from numerous massive stars in a star cluster is able to mix efficiently with the surrounding cold, turbulent ISM due to the fractal nature of the surrounding turbulent medium. This effect then leads to an efficient mixing layer between the hot and cold gas that rapidly cools via radiation and the dynamics of the wind-driven bubble is primarily momentum-conserving since the majority of the injected kinetic energy is lost via radiation. 

In contrast to the results of \citet{Lancaster2021a}, we find that the wind bubbles blown by individual high-mass stars do not experience efficient mixing because the magnetic field in the swept-up cold gas in the shell has a confining and stabilizing effect, regardless of the surrounding turbulent structure of the ISM. This is because the magnetic field in the ISM suppresses the growth of Rayleigh-Taylor and Kelvin-Helmholtz instabilities, which occur along the contact discontinuity between the shocked wind and the shocked ISM, which has been seen in previous numerical studies of wind feedback of a high-mass star moving in an ambient medium \citep{vanMarle2014a, Meyer2017a}. The suppression of instabilities therefore makes mixing between the hot and cold gas inefficient, thereby allowing for the thermal pressure in the wind bubble to control the dynamics of the expanding swept-up shells until the bubbles become pressure confined. Since our simulations only include ideal-MHD effects, the magnetic field in the shell will increase to conserve magnetic flux as the shell expands thereby further inhibiting the development of instabilities in the expanding shell \citep{Krumholz2007a}. Therefore, we find that, at least in the context of wind feedback from individual high-mass stars, the wind bubbles should be energy-driven rather than momentum-driven.  Non-ideal magnetic effects, such as ambipolar diffusion which is not explored in this work, may reduce the magnitude of the swept-up magnetic field thereby allowing for more efficient mixing and subsequent cooling at the bubble interface \citep{Fierlinger2016a}. 


\subsection{Other Bubble Driving Mechanisms}
\label{sec:sources}
Our results demonstrate that the kinetic energy injected by radiatively-driven wind feedback from intermediate-mass stars is ineffective at driving $\sim$pc scale bubbles that are observed around young, intermediate-mass stars. 
However, we do find that wind feedback from high-mass early type B stars would be able to produce bubbles with sizes of up to a pc. We note that we do not include self-consistent star formation and instead place the stars in a pristine cool molecular cloud with a high gas density. This is partially motivated by the idea that some of the observed B-type stars may not have formed in the observed region but have instead traveled there from a nearby, now-dispersed cloud \citep[e.g.,][]{Arce2011a}. However, this means that  there is no imprint from stellar feedback that would have been produced if the star had formed in the cloud. The effect of this earlier protostellar feedback during the star formation process would have evacuated gas near the stars and thus likely would have allowed the formation of larger wind bubbles. Regardless, our results suggest that main-sequence winds alone from intermediate-mass stars are likely not responsible for the observed shells and bubbles in star-forming regions. Here we discuss other possible shell driving mechanisms.

\subsubsection{Photoevaporative Disk Winds}
As PMS stars contract to the main-sequence they are usually surrounded by a circumstellar disk. This disk will eventually be dispersed by photoevaporative (PE) disk winds that are driven by irradiation from high-energy photons (i.e., X-rays, EUV, and FUV) emitted from the star. Photoevaporative disk dispersal is dominated by the stellar FUV radiation field for stars $\gtrsim 3 \; \rm \msun$ and these disk winds will be broader than the highly-collimated magnetically-launched protostellar jets launched near the star \citep{Gorti2009a, Owen2012a, Komaki2021a, Kunitomo2021a}. This is because FUV photoevaporation predominantly removes less bound gas from the outer disk further broadening the wind outflow \citep{Gorti2009a}. Additionally, \citet{Gorti2009a} find that the lifetimes of disks around intermediate-mass stars range from $\sim 10^5-10^6$ yr, with disk lifetimes decreasing for increasing stellar mass due to their high EUV and FUV fields. These  lifetime estimates are roughly in agreement with the ages of the shells observed in Orion \citep{Feddersen2018a}.

By performing radiation-hydrodynamics simulations of PE disk dispersal around low- and intermediate-mass stars \citet{Komaki2021a} found that the mass-loss rate increases with stellar mass and can be approximated as $\dot{M}_{\rm w} \approx 7.3 \times 10^{-9} \left(M_{\rm \star}/\msun \right)^2 \; \rm{\msun \; yr^{-1}}$, which ranges from $\approx 6.6 \times 10^{-8} - 4.7\times 10^{-7}  \; \rm{\msun \; yr^{-1}}$ for $M_{\rm \star} = 3 \; \msun - 8 \; \msun$, yielding mass-loss rates 2-3 orders of magnitude larger than the radiatively-driven mass-loss rates expected for intermediate-mass stars. Numerical and observational studies find that disk winds have velocities of $\lesssim 30 \; \rm km \; s^{-1}$ because magnetic stresses remove angular momentum from the wind as it is ejected from the disk \citep[e.g., ][]{Ercolano2017a, Gudel2018a, Gressel2020a}. With these values, the disk wind momenta are less than those inferred from observations of the shells observed in Perseus, Taurus, and Orion. Therefore, we conclude that although the mass-loss rates associated with disk winds are in better agreement with the mass-loss rates inferred from observations they are likely not responsible for the observed shells in star-forming regions. 



\subsubsection{Multiple Outflows}
During their formation, stars launch collimated bipolar outflows that entrain molecular material as they propagate away from the star \citep[e.g.,][]{Lada1985a, Bally2016a}. These outflows are magnetically launched due to the interaction between the magnetic fields and rotating star-disk system, and have two components: a $\sim100$ km/s collimated jet and a slower, radially distributed wide-angle disk wind that broadens with protostellar age \citep{Arce2006a, ZhangArce2019a}. The resulting mass-loss rates are $\sim$10-30\% of the accretion rate, and the gas is launched at a fraction of the star's escape speed \citep[e.g.,][]{Shu1988a, Pelletier1992a, Matzner2000a}. During the primary accretion phase, the accretion rate is approximately $10^{-5}-10^{-4} \; \rm{\msun \; yr^{-1}}$ and therefore the outflow mass-loss rates for intermediate-mass stars are orders of magnitude higher than those for radiatively-driven winds, in agreement with the estimated mass-loss rates from observed shells in Perseus and Orion \citep{McKee2003a, Rosen2012a, Arce2011a, Offner2017a, Feddersen2018a}. 

\citet{Offner2015a} demonstrated that the momentum input from collimated outflows is comparable to that needed to explain the dynamics of the expanding shells in Perseus, but they note that individual collimated outflows are unlikely to produce roughly spherical shells. However, it remains unclear if these shells are driven by only one source or by multiple sources. For example,  many of the identified shells in Orion may contain and be created by multiple PMS stars \citep{Feddersen2018a}. Similarly, several pc-scale CO shells in Ophiuchus are located near small, young clusters and may have been produced by the combined outflows from multiple sources \citep{Nakamura2012a}. The integrated momentum from these outflows may drive a roughly spherical structure if the  molecular outflows are misaligned, overlap, or change direction with time \citep{Lee2016a, LeeHull2017a, Avison2021a}. As these collimated outflows propagate away from the stars they will entrain magnetized molecular material and the influence of magnetic fields will broaden the outflows and produce a larger opening angle \citep{Rosen2020b}. 
Likewise, 
many of the shells in Taurus are ``broken" (i.e., not fully spherical) suggesting that feedback from the sources does not necessarily have to be isotropic \citep{Lili2015a,Xu2020a}. Therefore, we suggest that the combined outflows of small clusters of stars may drive these roughly spherical and broken shells in star-forming environments. 

In addition, collimated outflows are accretion powered and therefore accretion variability from individual or multiple sources might lead to high enough mass-loss rates to produce the observed shells in Perseus and Orion \citep{Offner2015a, Feddersen2018a}. In this scenario, young low- and intermediate-mass stars show variability in accretion as they contract to the main-sequence and therefore may experience periodic wind enhancements due to enhanced magnetic stellar activity or accretion-induced outflows \citep[e.g., see review by][]{Hartmann2016a}. \citet{Feddersen2018a} estimate that the shells in Orion have ages of $\sim0.1$ Myr and therefore such winds would have to be short-lived. 

\section{Conclusions}
\label{sec:conclusions}
In this work, we performed a series of 3D MHD simulations following the impact of radiatively-driven stellar wind feedback from young intermediate- and high-mass stars within a turbulent molecular cloud to determine if the momentum and energy injected by main-sequence stellar winds can produce the $\sim$pc scale bubbles observed via CO molecular line emission in star forming environments. This is one of the first studies that followed the kinetic energy injection from stellar winds for intermediate- and high-mass stars into a turbulent and magnetized molecular cloud rather than assume that the kinetic energy is instantly lost via radiation (i.e., momentum-driven feedback). This has allowed us to study how the shock-heating of the fast, stellar wind material can generate fast, hot, and low-density gas that cools via adiabatic expansion (i.e., energy-driven feedback), thereby suffering little radiative losses and drives the shell expansion of the swept-up material as the hot bubbles expand. We reach the following conclusions.

\begin{enumerate}
\item We performed two simulations, the first of which assumes an isothermal equation of state (\isoRun) for the gas, which assumes the kinetic energy from winds is  instantly lost via radiative cooling, and the second assumes an adiabatic equation of state (\adiaRun) that properly tracks the shock-heating of the kinetic energy injected by winds. We find that energy-driven feedback from main-sequence stellar winds launched by high-mass stars ($M_{\rm \star} \gtrsim 10 \rm \; \msun$) drive $\sim$pc scale bubbles but that such feedback is not sufficient at driving similar bubbles around intermediate-mass stars.
\item Regardless if we consider energy- or momentum-driven feedback, we find that the wind-driven bubbles eventually become pressure confined due to the turbulent and magnetic pressures in the surrounding ISM.
\item We find that, regardless of the equation of state modeled, wind feedback from winds drives local turbulence in molecular clouds and that energy-driven feedback from winds predominately drives low-density solenoidal turbulence within the clouds.
\item We see a larger increase for solenoidal turbulence driving in run \adiaRun\  as compared to \isoRun\ because the overall effect of adiabatic expansion generates low-density, fast flowing gas that is produced by the shock heating of stellar winds. Additionally, energy-driven wind feedback drives larger scale bubbles thereby increasing the volume filling factor of the hot gas contained within the bubbles.
\item We find that the adiabatic bubbles produced by wind feedback from individual high-mass stars do not cool efficiently via radiation and that the swept-up magnetic fields in the bubble shells suppress the development of Kelvin-Helmholtz and Rayleigh-Taylor instabilities at the hot-cold interface between the expanding bubbles and surrounding cold ISM material. 
\item We also find that minor instabilities develop at the shells that surround the adiabatic wind bubbles blown by winds from high-mass stars and these instabilities are likely attributed to the magnetic kink instability. Furthermore, their subsequent growth is suppressed due to the constant energy and momentum injection by stellar winds.
\item Given that the energy and momentum feedback main-sequence winds from intermediate-mass stars is not sufficient to drive these observed bubbles, we suggest that other bubble-driving mechanisms, such as collimated protostellar outflows from multiple stars or from variable accretion-driven winds, may be responsible for driving shells in star-forming environments. 
\end{enumerate}

\software{\orion\ \citep{Li2012a}, \textsc{yt} \citep{Turk2011a}, \textsc{unyt} \citep{Goldbaum2018a}, \chianti\ \citep{Dere1997a}}

\subsection*{Acknowledgements}
A.L.R. acknowledges support from NASA through Einstein Postdoctoral Fellowship grant number PF7-180166 awarded by the \textit{Chandra} X-ray Center, which is operated by the Smithsonian Astrophysical Observatory for NASA under contract NAS8-03060; and support from Harvard University through the ITC Post-doctoral Fellowship. S.S.R.O acknowledges support for this work from NSF Career grant 1748571 and NSF AST-1812747. M.M.F. acknowledges support from the NSF Graduate Research Fellowship Program. L.A.L. is supported by a Cottrell Scholar Award from the Research Corporation of Science Advancement. The simulations were run on the NASA supercomputer Pleiades located at NASA Ames. We use the \textsc{yt} package \citep{Turk2011a} to produce all the figures and quantitative analysis.

\bibliographystyle{apj}
\bibliography{refsMC}
\end{document}